\pdfoutput=1
\documentclass[12pt,a4paper,titlepage]{article}
\usepackage[utf8]{inputenc}
\usepackage[top=50pt,bottom=50pt,left=68pt,right=66pt]{geometry}
\usepackage{amsmath}
\usepackage{booktabs}
\usepackage{amssymb}
\usepackage[title]{appendix}
\usepackage{mathrsfs}
\usepackage{float}
\usepackage{multirow}
\usepackage{graphicx,caption,subcaption}
\usepackage[space]{grffile}

\begin{document}
\renewcommand{\arraystretch}{1.3}

\makeatletter
\def\@hangfrom#1{\setbox\@tempboxa\hbox{{#1}}%
      \hangindent 0pt
      \noindent\box\@tempboxa}
\makeatother


\def\un#1{\relax\ifmmode\@@underline#1\else
        $\@@underline{\hbox{#1}}$\relax\fi}


\let\under=\unt                 
\let\ced=\ce                    
\let\du=\du                     
\let\um=\Hu                     
\let\sll=\lp                    
\let\Sll=\Lp                    
\let\slo=\os                    
\let\Slo=\Os                    
\let\tie=\ta                    
\let\br=\ub                     


\def\a{\alpha}
\def\b{\beta}
\def\c{\chi}
\def\d{\delta}
\def\e{\epsilon}
\def\f{\phi}
\def\g{\gamma}
\def\h{\eta}
\def\i{\iota}
\def\j{\psi}
\def\k{\kappa}
\def\l{\lambda}
\def\m{\mu}
\def\n{\nu}
\def\o{\omega}
\def\p{\pi}
\def\q{\theta}
\def\r{\rho}
\def\s{\sigma}
\def\t{\tau}
\def\u{\upsilon}
\def\x{\xi}
\def\z{\zeta}
\def\D{\Delta}
\def\F{\Phi}
\def\G{\Gamma}
\def\J{\Psi}
\def\L{\Lambda}
\def\O{\Omega}
\def\P{\Pi}
\def\Q{\Theta}
\def\S{\Sigma}
\def\U{\Upsilon}
\def\X{\Xi}


\def\ve{\varepsilon}
\def\vf{\varphi}
\def\vr{\varrho}
\def\vs{\varsigma}
\def\vq{\vartheta}


\def\ca{{\cal A}}
\def\cb{{\cal B}}
\def\cc{{\cal C}}
\def\cd{{\cal D}}
\def\ce{{\cal E}}
\def\cf{{\cal F}}
\def\cg{{\cal G}}
\def\ch{{\cal H}}
\def\ci{{\cal I}}
\def\cj{{\cal J}}
\def\ck{{\cal K}}
\def\cl{{\cal L}}
\def\cm{{\cal M}}
\def\cn{{\cal N}}
\def\co{{\cal O}}
\def\cp{{\cal P}}
\def\cq{{\cal Q}}
\def\car{{\cal R}}
\def\cs{{\cal S}}
\def\ct{{\cal T}}
\def\cu{{\cal U}}
\def\cv{{\cal V}}
\def\cw{{\cal W}}
\def\cx{{\cal X}}
\def\cy{{\cal Y}}
\def\cz{{\cal Z}}


\def\Sc#1{{\hbox{\sc #1}}}      
\def\Sf#1{{\hbox{\sf #1}}}      



\def\slpa{\slash{\pa}}                            
\def\slin{\SLLash{\in}}                                   
\def\bo{{\raise-.3ex\hbox{\large$\Box$}}}               
\def\cbo{\Sc [}                                         
\def\pa{\partial}                                       
\def\de{\nabla}                                         
\def\dell{\bigtriangledown}                             
\def\su{\sum}                                           
\def\pr{\prod}                                          
\def\iff{\leftrightarrow}                               
\def\conj{{\hbox{\large *}}}                            
\def\ltap{\raisebox{-.4ex}{\rlap{$\sim$}} \raisebox{.4ex}{$<$}}   
\def\gtap{\raisebox{-.4ex}{\rlap{$\sim$}} \raisebox{.4ex}{$>$}}   
\def\TH{{\raise.2ex\hbox{$\displaystyle \bigodot$}\mskip-4.7mu \llap H \;}}
\def\face{{\raise.2ex\hbox{$\displaystyle \bigodot$}\mskip-2.2mu \llap {$\ddot
        \smile$}}}                                      
\def\dg{\sp\dagger}                                     
\def\ddg{\sp\ddagger}                                   

\font\tenex=cmex10 scaled 1200


\def\sp#1{{}^{#1}}                              
\def\sb#1{{}_{#1}}                              
\def\oldsl#1{\rlap/#1}                          
\def\slash#1{\rlap{\hbox{$\mskip 1 mu /$}}#1}      
\def\Slash#1{\rlap{\hbox{$\mskip 3 mu /$}}#1}      
\def\SLash#1{\rlap{\hbox{$\mskip 4.5 mu /$}}#1}    
\def\SLLash#1{\rlap{\hbox{$\mskip 6 mu /$}}#1}      
\def\PMMM#1{\rlap{\hbox{$\mskip 2 mu | $}}#1}   %
\def\PMM#1{\rlap{\hbox{$\mskip 4 mu ~ \mid $}}#1}       %
\def\Tilde#1{\widetilde{#1}}                    
\def\Hat#1{\widehat{#1}}                        
\def\Bar#1{\overline{#1}}                       
\def\sbar#1{\stackrel{*}{\Bar{#1}}}             
\def\bra#1{\left\langle #1\right|}              
\def\ket#1{\left| #1\right\rangle}              
\def\VEV#1{\left\langle #1\right\rangle}        
\def\abs#1{\left| #1\right|}                    
\def\leftrightarrowfill{$\mathsurround=0pt \mathord\leftarrow \mkern-6mu
        \cleaders\hbox{$\mkern-2mu \mathord- \mkern-2mu$}\hfill
        \mkern-6mu \mathord\rightarrow$}
\def\dvec#1{\vbox{\ialign{##\crcr
        \leftrightarrowfill\crcr\noalign{\kern-1pt\nointerlineskip}
        $\hfil\displaystyle{#1}\hfil$\crcr}}}           
\def\dt#1{{\buildrel {\hbox{\LARGE .}} \over {#1}}}     
\def\dtt#1{{\buildrel \bullet \over {#1}}}              
\def\der#1{{\pa \over \pa {#1}}}                
\def\fder#1{{\d \over \d {#1}}}                 


\def\frac#1#2{{\textstyle{#1\over\vphantom2\smash{\raise.20ex
        \hbox{$\scriptstyle{#2}$}}}}}                   
\def\half{\frac12}                                        
\def\sfrac#1#2{{\vphantom1\smash{\lower.5ex\hbox{\small$#1$}}\over
        \vphantom1\smash{\raise.4ex\hbox{\small$#2$}}}} 
\def\bfrac#1#2{{\vphantom1\smash{\lower.5ex\hbox{$#1$}}\over
        \vphantom1\smash{\raise.3ex\hbox{$#2$}}}}       
\def\afrac#1#2{{\vphantom1\smash{\lower.5ex\hbox{$#1$}}\over#2}}    
\def\partder#1#2{{\partial #1\over\partial #2}}   
\def\parvar#1#2{{\d #1\over \d #2}}               
\def\secder#1#2#3{{\partial^2 #1\over\partial #2 \partial #3}}  
\def\on#1#2{\mathop{\null#2}\limits^{#1}}               
\def\bvec#1{\on\leftarrow{#1}}                  
\def\oover#1{\on\circ{#1}}                              

\def\[{\lfloor{\hskip 0.35pt}\!\!\!\lceil}
\def\]{\rfloor{\hskip 0.35pt}\!\!\!\rceil}
\def\Lag{{\cal L}}
\def\du#1#2{_{#1}{}^{#2}}
\def\ud#1#2{^{#1}{}_{#2}}
\def\dud#1#2#3{_{#1}{}^{#2}{}_{#3}}
\def\udu#1#2#3{^{#1}{}_{#2}{}^{#3}}
\def\calD{{\cal D}}
\def\calM{{\cal M}}

\def\szet{{${\scriptstyle \b}$}}
\def\ulA{{\un A}}
\def\ulM{{\underline M}}
\def\cdm{{\Sc D}_{--}}
\def\cdp{{\Sc D}_{++}}
\def\vTheta{\check\Theta}
\def\fracm#1#2{\hbox{\large{${\frac{{#1}}{{#2}}}$}}}
\def\ha{{\fracmm12}}
\def\tr{{\rm tr}}
\def\Tr{{\rm Tr}}
\def\itrema{$\ddot{\scriptstyle 1}$}
\def\ula{{\underline a}} \def\ulb{{\underline b}} \def\ulc{{\underline c}}
\def\uld{{\underline d}} \def\ule{{\underline e}} \def\ulf{{\underline f}}
\def\ulg{{\underline g}}
\def\items#1{\\ \item{[#1]}}
\def\ul{\underline}
\def\un{\underline}
\def\fracmm#1#2{{{#1}\over{#2}}}
\def\footnotew#1{\footnote{\hsize=6.5in {#1}}}
\def\low#1{{\raise -3pt\hbox{${\hskip 0.75pt}\!_{#1}$}}}

\def\Dot#1{\buildrel{_{_{\hskip 0.01in}\bullet}}\over{#1}}
\def\dt#1{\Dot{#1}}

\def\DDot#1{\buildrel{_{_{\hskip 0.01in}\bullet\bullet}}\over{#1}}
\def\ddt#1{\DDot{#1}}

\def\DDDot#1{\buildrel{_{_{\hskip 0.01in}\bullet\bullet\bullet}}\over{#1}}
\def\dddt#1{\DDDot{#1}}

\def\DDDDot#1{\buildrel{_{_{\hskip 
0.01in}\bullet\bullet\bullet\bullet}}\over{#1}}
\def\ddddt#1{\DDDDot{#1}}

\def\Tilde#1{{\widetilde{#1}}\hskip 0.015in}
\def\Hat#1{\widehat{#1}}


\newskip\humongous \humongous=0pt plus 1000pt minus 1000pt
\def\caja{\mathsurround=0pt}
\def\eqalign#1{\,\vcenter{\openup2\jot \caja
        \ialign{\strut \hfil$\displaystyle{##}$&$
        \displaystyle{{}##}$\hfil\crcr#1\crcr}}\,}
\newif\ifdtup
\def\panorama{\global\dtuptrue \openup2\jot \caja
        \everycr{\noalign{\ifdtup \global\dtupfalse
        \vskip-\lineskiplimit \vskip\normallineskiplimit
        \else \penalty\interdisplaylinepenalty \fi}}}
\def\li#1{\panorama \tabskip=\humongous                         
        \halign to\displaywidth{\hfil$\displaystyle{##}$
        \tabskip=0pt&$\displaystyle{{}##}$\hfil
        \tabskip=\humongous&\llap{$##$}\tabskip=0pt
        \crcr#1\crcr}}
\def\eqalignnotwo#1{\panorama \tabskip=\humongous
        \halign to\displaywidth{\hfil$\displaystyle{##}$
        \tabskip=0pt&$\displaystyle{{}##}$
        \tabskip=0pt&$\displaystyle{{}##}$\hfil
        \tabskip=\humongous&\llap{$##$}\tabskip=0pt
        \crcr#1\crcr}}


\def\eV{\,{\rm eV}}
\def\keV{\,{\rm keV}}
\def\MeV{\,{\rm MeV}}
\def\GeV{\,{\rm GeV}}
\def\TeV{\,{\rm TeV}}
\def\sv{\left<\sigma v\right>}
\def\({\left(}
\def\){\right)}
\def\cm{{\,\rm cm}}
\def\K{{\,\rm K}}
\def\kpc{{\,\rm kpc}}
\def\beq{\begin{equation}}
\def\eeq{\end{equation}}
\def\bea{\begin{eqnarray}}
\def\eea{\end{eqnarray}}


\newcommand{\be}{\begin{equation}}
\newcommand{\ee}{\end{equation}}
\newcommand{\nbe}{\begin{equation*}}
\newcommand{\nee}{\end{equation*}}

\newcommand{\fr}{\frac}
\newcommand{\lb}{\label}

\thispagestyle{empty}

{\hbox to\hsize{
\vbox{\noindent September 2022 \hfill IPMU22-0037}
\noindent  \hfill }

\noindent
\vskip2.0cm
\begin{center}

{\large\bf E-models of inflation and primordial black holes}

\vglue.3in

Daniel Frolovsky~${}^{a}$, Sergei V. Ketov~${}^{a,b,c,\#}$ and Sultan Saburov~${}^{a}$
\vglue.3in

${}^a$~Interdisciplinary Research Laboratory, Tomsk State University\\
36 Lenin Avenue, Tomsk 634050, Russia\\
${}^b$~Department of Physics, Tokyo Metropolitan University\\
1-1 Minami-ohsawa, Hachioji-shi, Tokyo 192-0397, Japan \\
${}^c$~Kavli Institute for the Physics and Mathematics of the Universe (WPI)
\\The University of Tokyo Institutes for Advanced Study,  \\ Kashiwa 277-8583, Japan\\
\vglue.1in

${}^{\#}$~ketov@tmu.ac.jp
\end{center}

\vglue.3in

\begin{center}
{\Large\bf Abstract}  
\end{center}

We propose and study the new (generalized) E-type $\alpha$-attractor models of inflation, in order to include formation
of primordial black holes (PBHs). The inflaton potential has a near-inflection point where slow-roll conditions are violated,
thus leading to large scalar perturbations collapsing to PBHs later.  An ultra-slow roll (short) phase exists between two (longer) phases of slow-roll inflation. We numerically investigate the phases of inflation, derive the power spectrum of scalar perturbations and calculate the PBHs masses. For certain values of the parameters, the asteroid-size PBHs can be formed with the masses of $10^{17}\div 10^{19}$ g, beyond the Hawking evaporation limit and in agreement with current CMB observations. Those PBHs are a candidate for (part of) dark matter in the present universe, while the gravitational waves induced by the PBHs formation may be detectable by the future space-based gravitational interferometers.

\newpage

\section{Introduction}

Measurements of the Cosmic Microwave Background (CMB) radiation 
by the Planck mission provide tight observational constraints on cosmological inflation in the early Universe \cite{Planck:2018jri,BICEP:2021xfz,Tristram:2021tvh}. Nevertheless, the simple Starobinsky model of inflation \cite{Starobinsky:1980te}, proposed the long time ago, is still consistent with the current precision measurements of the CMB spectral tilt $n_s$ of scalar perturbations
\cite{Planck:2018jri,BICEP:2021xfz,Tristram:2021tvh},
\be \lb{ns}
n_s = 0.9649 \pm 0.0042 \quad (68\%~{\rm C.L.})
\ee
The Starobinsky model also gives a prediction for the value of the CMB tensor-to-scalar ratio $r$ up to an uncertainty in the duration of inflation measured by the number of e-folds $N_e$ as
\be \lb{rstar}
r_{\rm S} \approx \fracmm{12}{N^2_e}~~,\quad {\rm where} \quad N_e = \int^{t_{\rm end}}_{t_{\rm initial}} H(t) dt~,
\ee
and $H(t)$ is the Hubble function. The current observational bound \cite{Planck:2018jri,BICEP:2021xfz,Tristram:2021tvh}
\be \lb{rplanck}
r < 0.036\quad (95\%~{\rm C.L.})
\ee
is already fulfilled for $N_e>20$, whereas the duration of inflation is expected at 
$N_e= 55\pm 10$. This estimate for $N_e$ comes from the predicted value of $n_s$ 
in the Starobinsky model via the Mukhanov-Chibisov formula \cite{Mukhanov:1981xt}
\be \lb{mc}
n_s \approx 1 - \fracmm{2}{N_e}~.
\ee

Equations (\ref{rstar}) and (\ref{mc}) for the tilts $r$ and $n_s$ show only the leading terms with respect to the inverse e-folds number $N_e$. Given higher precision of the $n_s$-measurements, the subleading terms may also be important. For example, in the case of
the Starobinsky model, one finds \cite{Kaneda:2010ut}
\begin{equation}\label{kkw}
	n_{s}=1-\fracmm{2}{N_{e}}+\fracmm{3\ln N_{e}}{2N_{e}^{2}} -\fracmm{4}{N^{2}_{e}} + {\cal O}\Bigg(\fracmm{\ln^{2}N_{e}}{N_{e}^{3}}\Bigg)~.
\end{equation}

The scalar potential of the canonical inflaton field $\phi$ in the Starobinsky model reads~\footnote{See e.g., Refs.~\cite{Ketov:2021fww,Ivanov:2021chn,Ketov:2022qwj} for details about the Starobinsky model, various extensions and applications. We do not reproduce here the standard equations describing background dynamics, perturbations and their power spectrum in single-field inflation, because they are well known and easily can be found in the literature.}
\be \lb{starp}
V_S(\phi) = \fracm{3}{4}M^2_{\rm Pl} M^2 ( 1 - y_S)^2~,
\ee
where we have introduced the dimensionless field 
\be \lb{y1}
y_S= \exp \left( -\sqrt{\fracmm{2}{3}} \fracmm{\phi}{M_{\rm Pl}}\right)
\ee
and the inflaton mass $M\sim 10^{-5}M_{\rm Pl}$, whose value is determined by the known CMB amplitude. The scale of inflation can be estimated by the Hubble function $H$ during slow-roll, which is related to the (unknown) 
tensor-to-scalar ratio $r$. As regards the Starobinsky inflation, the scale of inflation $H_S\sim M$ corresponds to super-high energy physics far beyond the electro-weak scale and not far from the GUT scale.

The flatness of the inflaton potential during slow roll is guaranteed by the smallness of $y_S$ during inflation. Therefore, the inflationary observables for CMB will be essentially the same (in the leading approximation with respect to $N^{-1}_e$ or $N^{-2}_e$) after a generalization of the scalar potential (\ref{starp}) to
\be \lb{vgen}
V_\z(\phi) = \fracm{3}{4}M^2_{\rm Pl} M^2 \left[ 1 - y_S + y_S^2\z(y_S) \right]^2~,
\ee
where $\z(y_S)$ is a function regular at $y_S=0$. Some generalizations of the Starobinsky model, like Eq.~(\ref{vgen}),  were studied in Ref.~\cite{Ivanov:2021chn}. In this paper, we take the inflaton potential to be a real function {\it squared} because it can always be minimally embedded into supergravity as a single-field inflationary model \cite{Ketov:2019toi}.

Another simple way of generalizing the Starobinsky model of inflation is given by the cosmological $\alpha-attractors$ \cite{Kallosh:2013hoa,Galante:2014ifa} that come in two families called E-models and T-models.
The E-models have the same scalar potential $V(y)$ as in Eq.~(\ref{starp}) but in terms of the new variable 
\be \lb{ya}
y = \exp \left( -\sqrt{\fracmm{2}{3\a}} \fracmm{\phi}{M_{\rm Pl}}\right)
\ee
that depends upon the parameter $\a>0$. The Starobinsky model corresponds to $\a=1$. The E-models lead to the same Eq.~(\ref{mc}) for the tilt $n_s$  but significantly change the tilt $r$ as
\be \lb{ralpha}
r_{\rm \a} \approx \fracmm{12\a}{N^2_e}~,
\ee
thus making this theoretical prediction more flexible against future measurements.

An opportunity of changing the inflaton potential by arbitrary function $\z(y)$ can be exploited in order to generate primordial black holes (PBHs) \cite{Novikov:1967tw,Hawking:1971ei} at smaller values of $\phi$ or,
 equivalently, at lower energy scales. Those energy scales (below the scale of inflation) are not tightly constrained by observations yet. Technically, the PBHs production  can be engineered by demanding a {\it near-inflection} point in the potential within the double inflation scenario with an ultra-slow-roll phase between two slow-roll regimes of inflation, leading to an enhancement of the power spectrum of scalar perturbations 
 \cite{Garcia-Bellido:2017mdw,Germani:2017bcs,Germani:2018jgr,Bhaumik:2019tvl}.~\footnote{See Ref.~\cite{Karam:2022nym} for a current review of PBHs formation in single-field inflationary models.} The PBHs born in the very early Universe are considered as a candidate for cold dark matter in the present Universe~\cite{Barrow:1992hq,Carr:2003bj,Sasaki:2018dmp,Carr:2020gox}. 
 
 A generalization of the Starobinsky model for PBHs formation was proposed and studied in Ref.~\cite{Frolovsky:2022ewg} by using a model
 very different from the $\alpha$-attractors.  As regards the generalized T-models of $\a$-attractors, the PBHs production was studied in Refs.~\cite{Dalianis:2018frf,Iacconi:2021ltm} for single-field inflation with the scalar potentials
\be \lb{tpot}
 V_T(\phi)=f^2\left(\tanh \fracmm{\phi/M_{\rm Pl}}{\sqrt{6\a}}\right)~,
 \ee
 where $f$ is a regular function.  In this paper, we propose and investigate the generalized E-models of inflation with a near-inflection point along similar lines.

 Our paper is organized as follows. In Sec.~2 we introduce our model and investigate its scalar potential. Section 3 is devoted to the slow-roll approximation during the first stage of inflation relevant to CMB. In Sec.~4 we give our results for the power spectrum of scalar perturbations and its enhancement leading to PBHs formation. Our conclusion is Sec.~5.

\section{The model}

Let us consider the following potential of the canonical inflaton $\phi$:
\be \lb{pot}
V(\phi) = \fracm{3}{4}(M_{\rm Pl} M)^2 \left[ 1 - y + y^2(\b - \g y)  \right]^2~,
\ee
with the dimensionless parameters $(\a,\b,\g)$, where the function $y(\phi)$ is given by
\be \lb{yfun}
y = \exp \left[ -\sqrt{\fracmm{2}{3\a}} \fracmm{(\phi+\phi_0)}{M_{\rm Pl}}\right]~.
\ee
Compared to Eqs.~(\ref{vgen}) and (\ref{ya}), we have Taylor-expanded the function $\z(y)$ up to a linear term, $\z(y)=\b - \g y$, and have shifted
the field $\phi$ by a constant $\phi_0$ in order to have a Minkowski minimum at $\phi=0$ with $V(0)=0$. Hence, the $\phi_0$ is fixed by other parameters. We do not give here an explicit formula for $\phi_0$ because it is not very illuminating.

Demanding the existence of a near-inflection point in the potential with a coordinate $\phi_{i}$ allows us to replace the parameters $(\b,\g)$ by the new dimensionless parameters $(\phi_{i},\x)$ as follows:
\be \lb{newpar}
\b = \fracmm{1}{1-\x^2} \exp \left[ \sqrt{\fracmm{2}{3\a}} \fracmm{(\phi_{i}+\phi_0)}{M_{\rm Pl}}\right]~,\quad
\g = \fracmm{1}{3(1-\x^2)} \exp \left[2\sqrt{\fracmm{2}{3\a}} \fracmm{(\phi_{i}+\phi_0)}{M_{\rm Pl}}\right]~.
\ee
The parameters $(\phi_{i},\x)$ have the clear meaning: when $\x=0$, the potential has the inflection point at 
$\phi=\phi_{i}$ only; when $0<\x\ll 1$, the potential has a local minimum $y_{\rm ext}^-$ on the right hand side of the inflection point $\phi_{i}$ and a local maximum $y_{\rm ext}^+$ on the left hand side of the inflection point $\phi_{i}$, while both extrema are equally separated from the inflection point,
\be \lb{extr2}  y^{\pm}_{\rm ext} = y_{i}\left( 1\pm \x\right) ~.
\ee
Equations (\ref{newpar}) and (\ref{extr2}) are easily derivable from considering extrema of the cubic 
polynomial inside the square brackets in (\ref{pot}), which leads to a quadratic equation ({\it cf.} Ref.~\cite{Iacconi:2021ltm}). The inverse
relations are given by
\be \lb{invnewpar}
\sqrt{\fracmm{2}{3\a}} \fracmm{(\phi_{i}+\phi_0)}{M_{\rm Pl}}=\ln \fracmm{3\g}{\b}~,\quad
\x^2=1-\fracmm{3\g}{\b^2}~.
\ee
In terms of the new parameters our scalar potential takes the form
\begin{equation}\label{pot2}
		\begin{split}
	V(\phi)=&\frac{3}{4}(MM_{\rm Pl})^{2}\Bigg\{1-\exp\Bigg[-\sqrt{\fracmm{2}{3\alpha}}\fracmm{(\phi+\phi_0)}{M_{\rm Pl}}\Bigg]+\fracmm{1}{1-\xi^{2}}\exp\Bigg[\sqrt{\fracmm{2}{3\alpha}}\fracmm{(\phi_{i}-2\phi_0-2\phi)}{M_{\rm Pl}}\Bigg]-{}\\	&-\fracmm{1}{3(1-\xi^{2})}\exp\Bigg[\sqrt{\fracmm{2}{3\alpha}}\fracmm{(2\phi_{i}-3\phi_0-3\phi)}{M_{\rm Pl}}\Bigg]\Bigg\}^2~~.
			\end{split}
\end{equation}

An example of the scalar potential leading to viable inflation and PBHs formation is given in Fig.~\ref{Figure1}. The potentials in the original
E-models of $\alpha$-attractors, arising in the case of $\beta=\gamma=\phi_0=0$, do not have a near-inflection point and thus do not lead to
PBHs formation. Our potential (\ref{pot2}) has the small bump, associated with the local maximum, and the small dip, associated with the local minimum, with both being close to the inflection point, similarly to the models  of Ref.~\cite{Mishra:2019pzq}.

\begin{figure}[h]
\begin{minipage}[h]{0.5\linewidth}
\center{\includegraphics[width=0.8\linewidth]{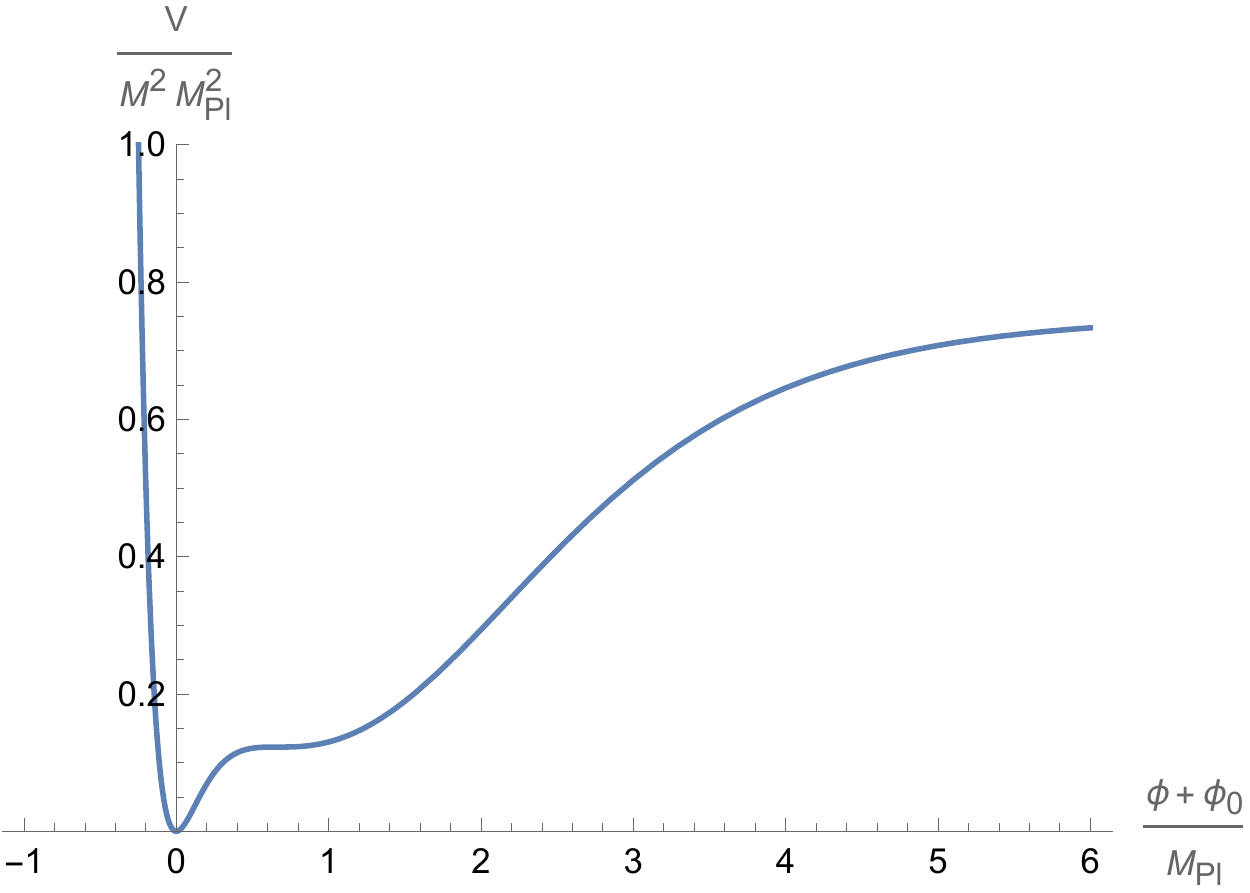} }
\end{minipage}
\hfill
\begin{minipage}[h]{0.5\linewidth}
\center{\includegraphics[width=0.8\linewidth]{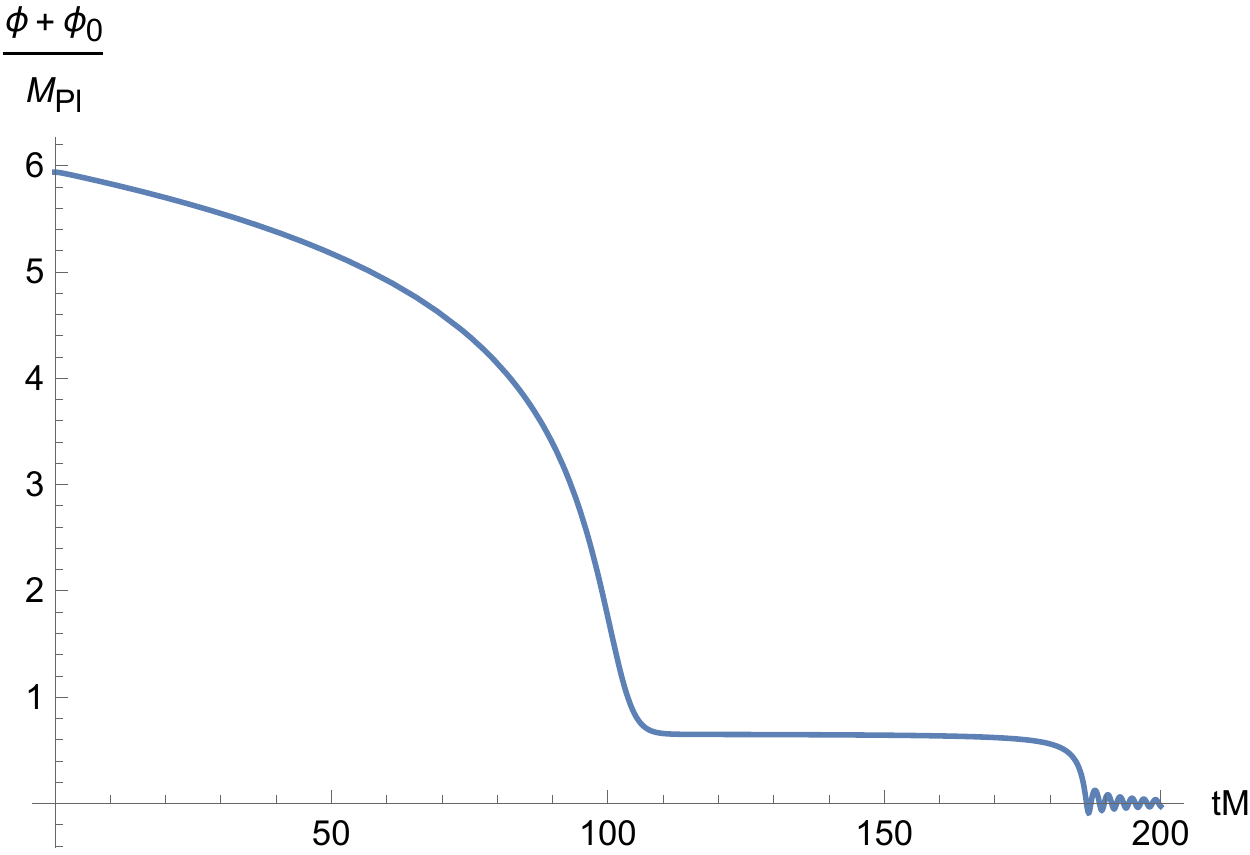} }
\end{minipage}
\caption{A profile of the scalar  potential (on the left) and the inflaton dynamics (on the right) for the parameters $\alpha=0.739$, 
$\phi_i+\phi_0=0.664M_{\rm Pl}$ and $\x=0.012$ with the vanishing initial velocity. The location of the inflection point is specified by  the value of
$(\phi_i+\phi_0)/M_{\rm Pl}$. }
\label{Figure1}
\end{figure}

\section{Slow-roll inflation}

Since the flatness of the scalar potential during inflation, the standard slow-roll approximation well describes  both the inflaton dynamics and the power spectrum of perturbations away from the inflection point and the end of inflation. We  employ the slow-roll approximation in order to calculate the observables relevant to CMB and estimate the power spectrum of scalar perturbations. It is known that the slow-roll approximation generically fails in the ultra-slow-roll (non-attractor) regime near an  inflection point \cite{Dalianis:2018frf,Iacconi:2021ltm}. Therefore, after having fixed our parameters in the slow-roll approximation, we numerically recalculate the power spectrum near the inflection point by using the  
Mukhanov-Sasaki (MS) equation \cite{Mukhanov:1985rz,Sasaki:1986hm} leading to a correct answer.

The (running) number of e-folds in the slow-roll approximation is given by
\begin{equation}\label{efoldsr}
	N_{e}=\int_{t}^{t_{end}}H(t)dt\approx\fracmm{1}{M_{\rm Pl}^{2}}\int_{\phi_{\rm end}}^{\phi}\fracmm{V(\phi)}{V'(\phi)}d\phi~,
\end{equation}
where the prime denotes differentiation with respect to the given argument. The integral can be taken analytically in the case of our potential (\ref{pot2}). We find
\begin{equation}\label{efapp}
	N_{e}(\phi)+N_0 \approx\fracmm{3\alpha}{4} \exp \left( \sqrt{\fracmm{2}{3\alpha}}\fracmm{(\phi+\phi_0)}{M_{\rm Pl}}\right) - \fracmm{3\sqrt{\alpha}}{4}\Bigg[1-2\exp\left( \sqrt{\fracmm{2}{3\alpha}}\fracmm{\phi_{i}}{M_{\rm Pl}}\right)\Bigg]\sqrt{\fracmm{2}{3}}\fracmm{(\phi+\phi_0)}{M_{\rm Pl}}~,
\end{equation}
where $N_0$ is an integration constant close to one. We ignore this constant for simplicity in what follows because it merely
shifts counting of $N_e$. The standard slow-roll parameters are given by
\begin{equation}\label{sreps}
\epsilon=\fracmm{M_{\rm Pl}^{2}}{2}\Bigg(\fracmm{V'(\phi)}{V(\phi)}\Bigg)^{2} 
= \fracmm{3\alpha}{4N^{2}_{e}}+ {\cal O}\Bigg(\fracmm{\ln^{2}N_{e}}{N_{e}^{3}}\Bigg)
\end{equation}
and
\begin{equation}\label{sreta}
\eta =M_{\rm Pl}^{2}\fracmm{V''(\phi)}{V(\phi)} = 
 -\fracmm{1}{N_{e}}+3\a(1-2e^{\sqrt{\frac{2}{3\alpha}}\frac{\phi_{i}}{M_{\rm Pl}}})\fracmm{\ln N_{e}}{4N_{e}^{2}}+\fracmm{3\a(1-2e^{\sqrt{\frac{2}{3\alpha}}\frac{\phi_{i}}{M_{\rm Pl}}})\ln\frac{4}{3\alpha}}{4N_{e}^{2}}+ {\cal O}\Bigg(\fracmm{\ln^{2}N_{e}}{N_{e}^{3}}\Bigg)~. 
\end{equation}
It yields
\be \lb{nsa}
n_{s}=1+2\eta-6\epsilon=1-\fracmm{2}{N_e}+a\fracmm{3\ln N_e}{2N_e^2} +\fracmm{b}{N^2_e} + 
{\cal O}\Bigg(\fracmm{\ln^{2}N_e}{N_e^3}\Bigg)~,
\ee
whose coefficients are given by
\be \lb{abc}
 a=\alpha\left(1-2e^{\sqrt{\frac{2}{3\alpha}}\frac{\phi_{i}}{M_{\rm Pl}}}\right) \quad {\rm and } \quad b=\fracmm{3\alpha}{2}
 \left[ \left(1-2e^{\sqrt{\frac{2}{3\alpha}}\frac{\phi_{i}}{M_{\rm Pl}}}\right)\ln\fracmm{4}{3\alpha}-3\right]~. 
\ee
Equation (\ref{sreps}) reproduces Eq.~(\ref{ralpha}) because $r=16\epsilon$. When choosing $\a=1$ and $\phi_{i}\to -\infty$,  Eq.~(\ref{kkw}) is also recovered up to a small correction $(=0.05)$ in the value of the coefficient $b$ due to our approximation.

\section{Power spectrum and PBH masses}

We numerically solve the inflaton equation of motion by using initial conditions with the vanishing initial velocities and then substitute the background solutions into the equations for perturbations. All our inflationary solutions are attractors (during slow roll) by construction. The initial  inflaton field value is fixed by a desired number of e-folds, see e.g., Refs.~\cite{Ketov:2021fww,Karam:2022nym} for details. 

A typical numerical solution to the Hubble function during double inflation is given on the left-hand-side of Fig.~\ref{image2}. Demanding a peak
in the power spectrum of scalar perturbations, required for PBHs production, we find the parameter $\alpha$ has to be restricted to the interval
between 0.5 and 0.9, whereas the parameter $\phi_i$ also has to be fixed, as is shown on the right-hand-side of Fig.~\ref{image2}. There is a short phase of ultra-slow-roll between the two stages of slow-roll inflation (corresponding to two plateaus), which leads to large perturbations in the power spectrum and PBHs production.

\begin{figure}[h]
\begin{minipage}[h]{0.5\linewidth}
\center{\includegraphics[width=0.8\linewidth]{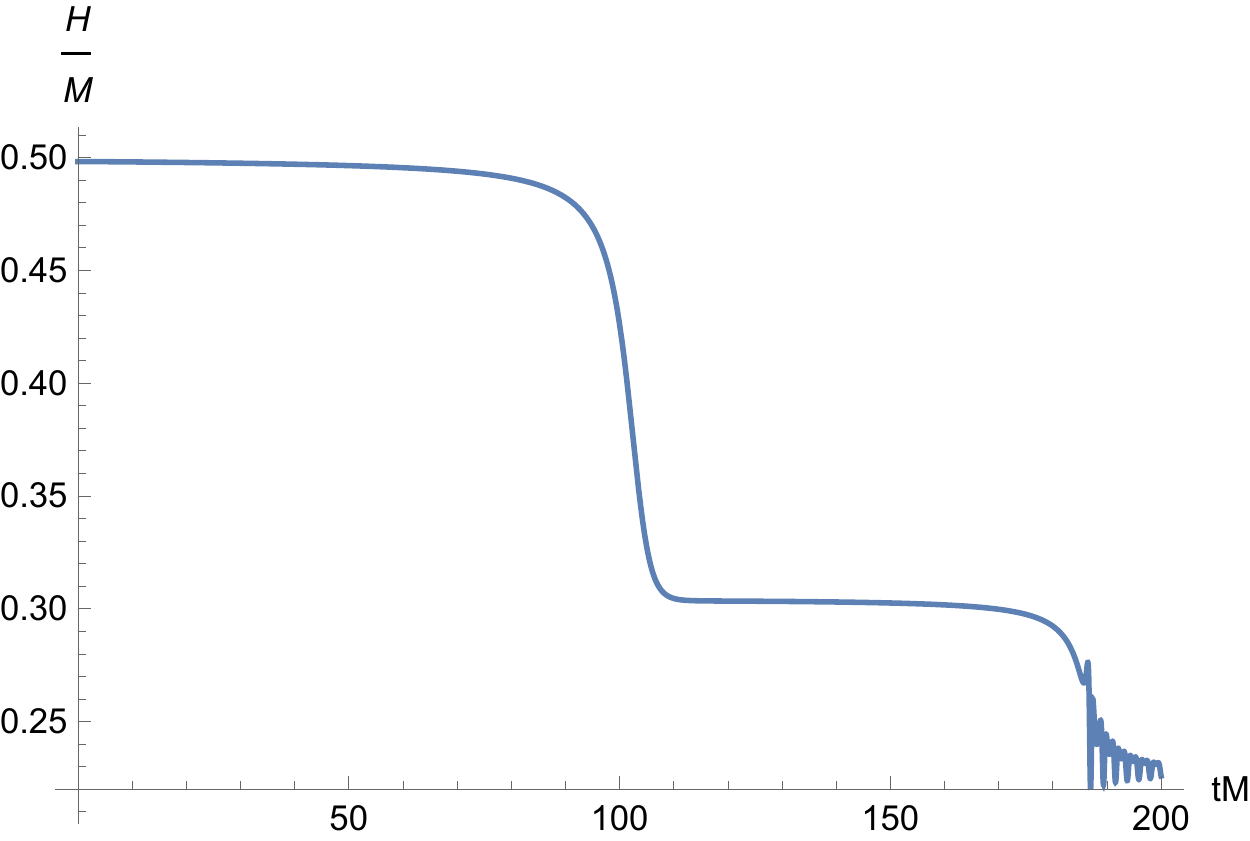} }
\end{minipage}
\hfill
\begin{minipage}[h]{0.5\linewidth}
\center{\includegraphics[width=0.8\linewidth]{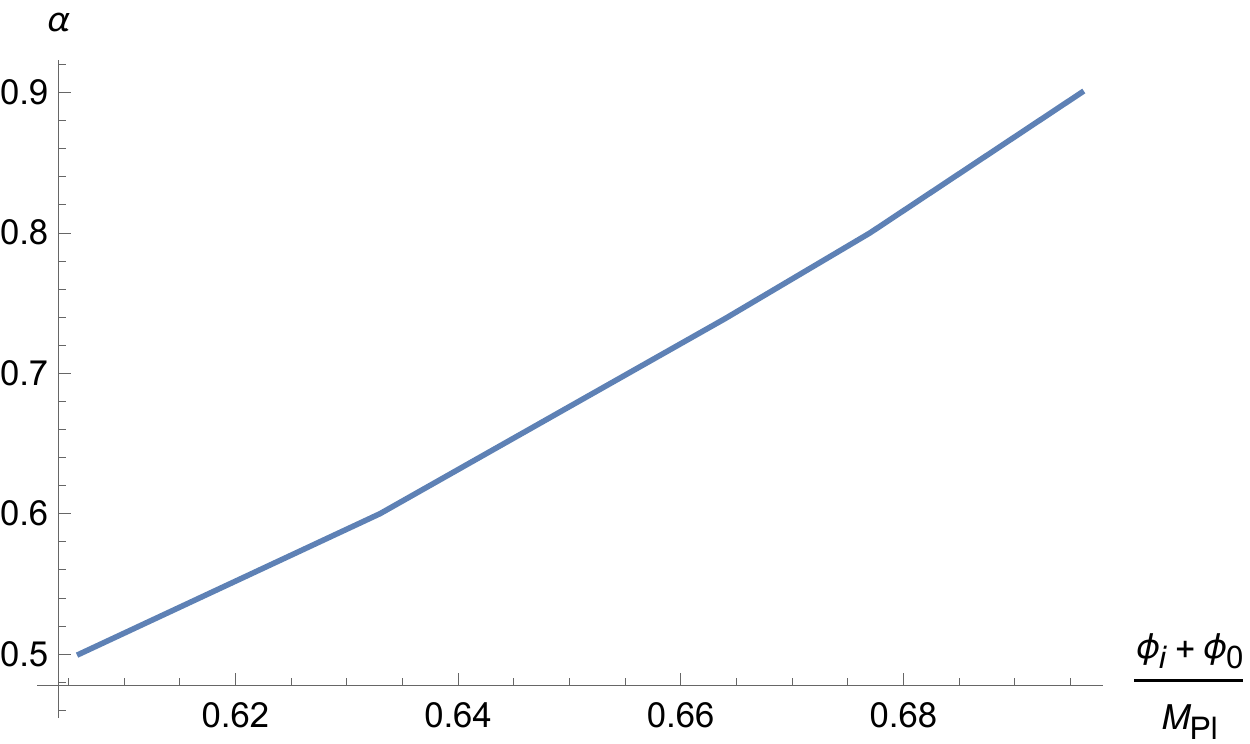} }
\end{minipage}
\caption{The Hubble function $H$ (on the left) and the relation between the parameters $\phi_i$ and $\alpha$ for the power spectrum enhancement and  PBHs production (on the right).}
\label{image2}
\end{figure}

The standard formula for the power spectrum of scalar perturbations in the slow-roll approximation
\cite{Garcia-Bellido:2017mdw}
\begin{equation} \lb{powersp}
	P_{R}=\fracmm{H^2}{8M^2_{\rm Pl}\pi^2\epsilon}
\end{equation}
is useful for analytic studies of the power spectrum and its dependence upon the parameters. However, it cannot be used in the ultra-slow-roll phase where the slow-roll conditions are violated. Instead, one should
use the MS equation \cite{Mukhanov:1985rz,Sasaki:1986hm}. We used both in our calculations in order to see a difference between the two methods.

The scalar tilt $n_s$ is related to the power spectrum by a relation $n_s=\fracmm{d \ln P_R}{d\ln k}$, where 
$k=aH=da/dt$ and $a(t)$ is the cosmic factor in the Friedman-Lemaitre-Robertson-Walker metric. Our results for the power spectrum are given in Fig.~\ref{image3} for a particular choice of the parameters. Our results are qualitatively similar for  other values of the parameters, see the right-hand-side of Fig.~\ref{image2}.

\begin{figure}[h]
\center{\includegraphics[width=0.6\linewidth]{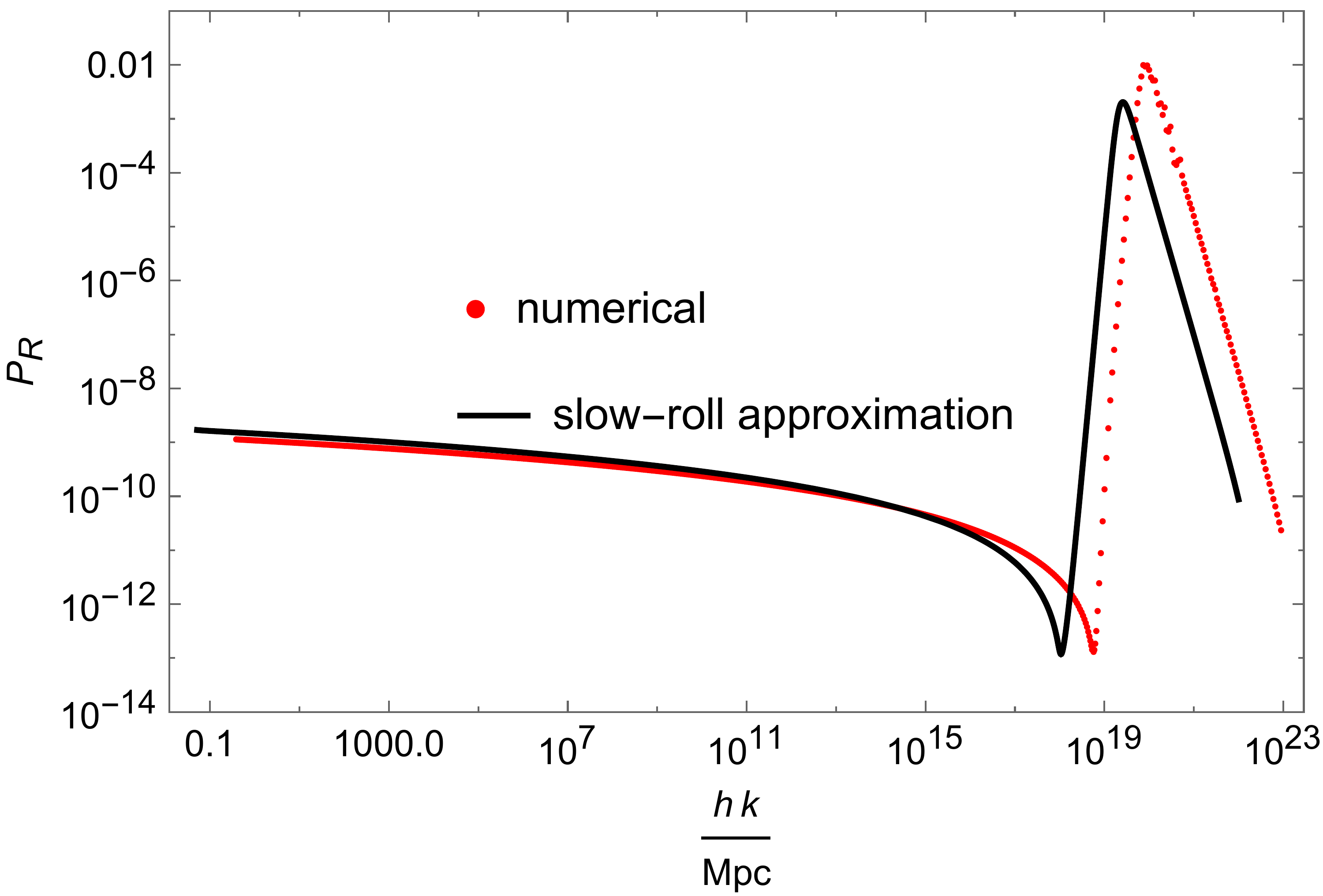}}
\caption{The power spectrum $P_{R}(k)$ of scalar perturbations from a numerical solution to the MS equation (in red) versus an analytic derivation from Eq.~(\ref{powersp}) in the slow-roll approximation (in black), with the same parameters as in Fig.~\ref{Figure1}. }
\label{image3}	
\end{figure}

As is clear from Fig.~\ref{image3}, the exact results based on the MS equation versus the slow-roll approximation increase the hight of the peak by one or two orders of magnitude, whereas the amplification of the peak versus the CMB spectrum (on the very left-hand-side of the power spectrum) is given by the seven orders of magnitude.

The PBHs masses can be estimated from the peaks as follows~\cite{Pi:2017gih}:
\begin{equation} \lb{pbhm}
M_{\rm PBH}\simeq \fracmm{M_{\rm Pl}^2}{H(t_{\rm peak})} \exp \left[2(N_{\rm total}-N_{\rm peak})+\int_{t_{\rm peak}}^{t_{\rm total} } \epsilon(t) H(t) dt    \right]~~.
\end{equation}
The right-hand-side of this equation is mainly sensitive to the value of $\D N=N_{\rm total}-N_{\rm peak}$, whereas the integral gives a sub-leading correction.  

Our findings are summarized in the Table below where we give the values of the CMB tilts $n_s$ and $r$ associated with the values of the parameters $\alpha$, $\phi_i$ and $\x$, together with the corresponding values of $\D N$ and PBHs masses $M_{\rm PBH}$ in our model.

\vglue.2in
\begin{table}[h]
\label{table}
\begin{center}
\begin{tabular}{| c | c | c | c | c | c | c | c | }
\hline
\phantom{@}$\mathbf{n_s}$\phantom{@} & \phantom{@}$\mathbf{r}$\phantom{@} & \phantom{@}\mbox{\boldmath{$\alpha$}} \phantom{@} & \phantom{@}\mbox{\boldmath{$\xi$}} \phantom{@} & \phantom{@}\mbox{\boldmath{$\phi_{\textbf{\text{i}}}$}}\phantom{@} &\phantom{@}\mbox{\boldmath{$\phi_{\textbf{\text{i}}}+\phi_{\textbf{\text{0}}}$}}\phantom{@} & \phantom{@}$\mathbf{\Delta N}$\phantom{@} & \phantom{@} $\mathbf{M_{\textbf{\text{PBH}}}}$ \phantom{@}\\ \hline
0.95452 & 0.00307 & 0.5  &  0.0102 & -\,0.334 & 0.606 & 15.08 & $1.06 \cdot 10^{19}$ g
\\ \hline
0.95491 & 0.00360 & 0.6 & 0.0106 & -\,0.455 & 0.633 & 15.35 & $1.04\cdot 10^{19}$ g
\\ \hline
0.95658 & 0.00409 & 0.739 & 0.0122 & -\,0.611 & 0.664 & 13.28 & $1.89 \cdot 10^{17}$ g \\ \hline
0.95672 & 0.00439 & 0.8 & 0.0115 & -\,0.671 & 0.677 & 13.96 & $7.75 \cdot 10^{17}$ g \\ \hline
0.95650 & 0.00496 & 0.9 &  0.0111 & -\,0.765  & 0.696 & 13.74 & $8.84 \cdot 10^{17}$ g  \\ \hline
\end{tabular}
\end{center}
\end{table}

The $n_s$ values below 0.9545 are certainly excluded by CMB observations, so we do not include our results for the lower values of $n_s$, see
Eq.~(\ref{ns}). The values of $n_s$ above 0.9565 are in good agreement with CMB observations at the 95\% C.L. The values of the tensor-to scalar ratio $r$ in the Table are well inside the current observational bound (\ref{rplanck}). We also found that lowering the value of the parameter  $\alpha$ leads to narrowing the peaks in the scalar perturbations spectrum. The PBHs masses are very sensitive to the value of $\D N$.

PBHs may be part of the present dark matter when the PBH masses are beyond the Hawking evaporation limit of $10^{15}$ g, which is required for survival of those PBHs in the present universe. However, consistency with the measured CMB value of $n_s$ restricts $\D N$ from above, as is clear from the Table.

\section{Conclusion}

Our approach is this paper is phenomenological and classical. However, it is not excluded that our deformations of the E-models of inflation proposed
in this paper could appear as quantum corrections from a more fundamental theory of quantum gravity.

We modified the scalar potential of the single-field E-models of $\a$-attractors in order to allow PBHs formation in those models at lower scales, while keeping success in the theoretical description of large-single-field inflation in agreement with CMB  measurements. We  found that efficient PBHs production consistent with CMB measurements restricts the $\alpha$ parameter to approximately $0.7\pm 0.2$ and leads to the asteroid-size PBHs with masses of the order $10^{17}\div 10^{19}$ g. The masses of the PBHs formed in the very early universe may grow further with time via accretion and mergers.

A similar approach was realized in the T-models of $\a$-attractors~\cite{Dalianis:2018frf,Iacconi:2021ltm}. In 
terms of pole inflation \cite{Galante:2014ifa} with a non-canonical inflaton field having just a mass term, the kinetic terms in the E-models have a pole  of order two and exhibit the $SL(2,R)$ symmetry, whereas the kinetic terms in the T-models also have a pole of order two but with the $SU(1,1)$ symmetry. Since those symmetries are equivalent, the main predictions of the standard E- and T-models for inflation are essentially the same. The
generalized E-models of inflation proposed in this paper simultaneously describe viable inflation and PBHs formation.

The next generation of CMB measurements will probe deeper regions of parameter space, leading to a discrimination among currently viable models of inflation, which may falsify the Starobinsky model in particular. The $\a$-attractors add more flexibility on the theoretical side, as regards the tensor-to-scalar ratio. We demonstrated that certain deformations of the scalars potentials in the E-models can also lead to efficient PBHs production capable to describe a whole (or part of) dark matter in the present universe.

We tuned the parameters of our model in order to overcome the Hawking radiation bound $10^{15}$ g for the PBHs masses, so that those PBHs may contribute to the current dark matter. Remarkably, the PBHs with the masses between $10^{17}$ g and $10^{19}$ g belong to the current observational mass window where those PBHs may constitute the whole dark matter \cite{Sasaki:2018dmp,Carr:2020gox}.
With lower PBHs masses we found no strong constraints on the parameters, but those PBHs  should all evaporate until now. Still, those PBHs may have dominated the early universe, while their remnants could form dark matter at present.

The PBHs formation in the very early universe should lead to a stochastic background of gravitational waves (GW) at present 
\cite{Saito:2008jc}.~\footnote{See e.g., Ref.~\cite{Domenech:2021ztg} for a current review.}
The frequency of those GW can be estimated as
\be \lb{gwf}
f_{\rm GW} \approx \left( \fracmm{ M_{\rm PBH}}{10^{16}~{\rm g}}\right)^{-1/2} ~~{\rm Hz}~.
\ee
It was argued in the literature \cite{Garcia-Bellido:2017aan,Cai:2018dig,Bartolo:2018evs} that those GW may be detectable by the future space-based gravitational interferometers such as LISA \cite{LISA}, TAIJI \cite{TAIJI}, TianQin \cite{TQ} and DECIGO \cite{DEC}.

\section*{Acknowledgements}

This work was supported by Tomsk State University under the development program Priority-2030.  SVK was also supported by Tokyo Metropolitan University, the Japanese Society for Promotion of Science under the grant No.~22K03624, and the World Premier International Research Center Initiative (MEXT, Japan).

\bibliography{Bibliography}{}
\bibliographystyle{utphys}

\end{document}